\documentstyle[prl,aps,multicol]{revtex}
\renewcommand{\narrowtext}{\begin{multicols}{2} \global\columnwidth20.5pc}
\renewcommand{\widetext}{\end{multicols} \global\columnwidth42.5pc}

\begin{document}
\newcommand{\be}{\begin{equation}}
\newcommand{\ee}{\end{equation}}
\newcommand{\bary}{\begin{eqnarray}}
\newcommand{\eary}{\end{eqnarray}}
\newcommand{\ea}{E_{\alpha}}
\newcommand{\eb}{E_{\beta}}
\newcommand{\eo}{\epsilon_0}
\newcommand{\ek}{E_{k}}
\newcommand{\eea}{\epsilon_\alpha}
\newcommand{\eeb}{\epsilon_\beta}
\newcommand{\rv}{r_{v}}
\newcommand{\ua}{u_{\alpha}}
\newcommand{\va}{v_{\alpha}}
\newcommand{\ub}{u_\beta}
\newcommand{\vb}{v_\beta}
\newcommand{\ka}{k_\alpha}
\newcommand{\kb}{k_\beta}
\newcommand{\kf}{k_f}
\newcommand{\kr}{k_r}
\newcommand{\mua}{\mu_\alpha}
\newcommand{\mub}{\mu_\beta}
\newcommand{\Aa}{A_\alpha}
\newcommand{\Ba}{B_\alpha}
\newcommand{\Ab}{A_\beta}
\newcommand{\Bb}{B_\beta}
\newcommand{\Del}{\Delta_\infty}
\newcommand{\na}{n_\alpha}
\newcommand{\sa}{s_\alpha}
\newcommand{\kp}{k_{+}}
\newcommand{\km}{k_{-}}
\title{Effective Mass of a Vortex in a Clean Superconductor}
\author{J. H. Han$^1$, P. Ao$^2$ and X.-M. Zhu$^2$}
\address{Asia Pacific Center for Theoretical Physics, 207-43
              Cheongryangri-dong Dongdaemun-gu Seoul 130-012, Korea$^{1}$
  \\ and \\
Department of  Physics, Ume\aa{\ }
University, S-901 87
Ume\aa, Sweden$^2$}
\maketitle
\draft
\begin{abstract}
We calculate the effective mass of a single quantized vortex in the
BCS superconductor at finite temperature. Using the effective action for a
vortex, we arrive at
the mass formula as the integral of the spectral function
$J(\omega)/\omega^3$ over frequency. 
The spectral function is given in terms of the 
transition elements of the gradient of the Hamiltonian
between Bogoliubov-deGennes eigenstates. We show that core-core,
and core-extended transitions yield the vortex mass, near $T\!=\!0$,
of order of electron mass displaced by the normal core.
The extended-extended states
contributions are linearly divergent with the low frequency cutoff
$\omega_c$, in accordance with the Ohmic character of the
spectral function at low energies. We argue that the mass and friction
are closely related in this system and arise from the same
mechanism - interaction with the surrounding fermionic degrees of freedom. 
\end{abstract}
\narrowtext

The dynamics of an individual vortex or their array is an
integral part of our understanding of the superfluid matter. In
superconductors there have been phenomenological approaches to
write down a classical equation of motion of a vortex\cite{eom}. 
This ``effective" equation will in general rely on the information of 
the mass, along with the longitudinal and transverse forces that act on
the vortex. 

The vortex mass also enters in an important way in the tunneling
process\cite{tunnel,cl}. Estimates of the mass based on Ginzburg-Landau
functional gives a mass of order $m\kf$ per unit vortex
length\cite{phenom}, while more recent work employing microscopic BCS-type
analysis predicts a much larger mass $\sim m\kf (\kf
\xi_0)^2$\cite{mass,simanek}. The authors of Refs.\ \cite{mass,simanek}
focused on the core quasiparticle contribution to mass, and were limited to
zero temperature. We use in this work the effective action 
approach \cite{simanek,az}
%
%
to go beyond those limitations.

A quantized vortex is a phase singularity accompanied by a change in the
order parameter density near the center. A closer scrutiny using the
Bogoliubov-de Gennes (BdG) equation reveals that there are discrete energy
levels inside the vortex core\cite{deGennes}, in addition to a continuous
spectrum with energy $|E|$ larger than the gap whose wavefunction extends
beyond the core region. Both states are perturbed once the vortex is set
in motion, by a perturbing Hamiltonian equal to the inner product of the
fermion current with the vortex velocity. Virtual transitions are induced
among core and extended states and lead to an energy shift proportional to
the square of the vortex velocity. Dividing by the squared velocity gives
a mass in agreement with those of Refs.\ \cite{mass,simanek}.

Here we proceed via an effective action in the Matsubara representation. 
By integrating out the fermions from the BCS action and expanding the
remaining degrees of freedom up to second order in the vortex coordinate
one can obtain the effective action of a single vortex\cite{simanek,az}.
It is possible to {\it derive} the transverse as well as the longitudinal
forces on a vortex from this action\cite{az}. Likewise we derive below a
formal expression of the vortex mass valid at finite temperature. It
agrees with the Hamiltonian perturbation theory at $T=0$. 
Since core and extended states have non-vanishing overlap in the core
region one expects core-core, core-extended and extended-extended
transitions to contribute to the mass unlike previous theories which only
considered core-core processes. Using approximate forms for eigenstates we
explicitly compute the transition elements and the mass.  
At low temperature the mass is comparable to that of electrons displaced
by the hollow core of the vortex. 

The effective action by itself is non-local in time, hence dissipative,
and the concept of mass only arises as a result of the local-time
approximation.  We have deduced both the mass and friction of the vortex
from a single framework, with the relative significance of inertia and
dissipation largely determined by the motion itself. In this work we will
focus on the pure system with no disorder. 

The effective action of a single vortex, characterized by the
two-dimensional coordinate $r_v (\tau)$\cite{comment} in the BCS
superconductor,
is given by\cite{simanek,az}
\bary
S_{eff}&=&
{1\over 2\pi}\int_0^\beta \! d\tau \int_{-\infty}^{\infty}\!
d\tau^\prime |r_v (\tau)\!\!-\!\!r_v (\tau')|^2 \nonumber \\
 &\times& \int_0^\infty d\omega
e^{-\omega |\tau\!-\!\tau^\prime |}J(\omega).
\label{eq:eff_action}
\eary
In the theory of dissipative quantum systems $J(\omega)$ is known as
the spectral function\cite{cl}:
\be
J(\omega)\!\!=\!\!{\pi\over 4}\!\sum_{\alpha\beta}\delta
(\omega\!\!-\!\!|\ea\!\!-\!\!\eb|)
|f(\ea)\!\!-\!\!f(\eb)| \left|\langle \alpha|
\nabla_v H_0 |\beta\rangle\right|^2.
\label{eq:spectralJ}
\ee
It is the gradient of the Hamiltonian operator that governs the
longitudinal dynamics of the vortex. The transverse motion of the vortex,
which is of course crucial for the overall dynamics, is
not explicitly considered in this work. The longitudinal
action shown in Eq.\ (\ref{eq:eff_action})  is responsible for the mass
and also dissipation of the vortex.  The states $\alpha, \beta$
are the eigenstates of the Nambu Hamiltonian $H_0$ with the energy $\ea$
and $\eb$ respectively.

By assuming $r_v (\tau)\!-\!r_v (\tau')\!\approx\!(\tau\!-\!\tau' )
\dot{r}_v (\tau)$, one can rewrite the action as
\be
S_{eff}\approx  {1\over2}\int_0^\beta d\tau
\left\{\int_0^\infty d\omega {4J(\omega)
 \over{\pi\omega^3}}\right\}
\left({{d\rv}\over d\tau}\right)^2.
\label{eq:local_action}
\ee
The vortex mass follows as
\bary
M_v &=&
\sum_{\alpha\beta}\left| {{f(\ea)-f(\eb)} \over
 {(\ea\!-\!\eb)^3}}\right|
 |\langle\alpha| \nabla_v H_0|\beta\rangle|^2
\label{eq:m_eff1} \\
 &=&\sum_{\alpha\beta}
 \left| {{f(\ea)-f(\eb)} \over
 {\ea\!-\!\eb}}\right|
|\langle \alpha|\nabla_v |\beta\rangle |^2.
\label{eq:m_eff2}
\eary
for a homogeneous medium, a result known in  dissipative quantum 
mechanics\cite{ambegaokar}.

The eigenstates of the BdG equation is written in the general
form\cite{deGennes}
\be
\psi_\alpha (x,y,z) = {1\over\sqrt{L}}e^{ik_z z}\left(
\begin{array}{c}
  e^{i(\mu\!-\!{1\over 2})\theta}u_\alpha (r) \\
  e^{i(\mu\!+\!{1\over 2})\theta}v_\alpha (r)
\end{array} \right)
\ee
in the presence of a gap function
$\Delta(\vec{r})=\Delta(r)e^{-i\theta}$
for a singly-quantized vortex centered at the origin. The radial functions
are obtained by solving the coupled differential equation
\bary
 r^2 u^{\prime\prime}+ru^\prime \!+\![(k_r^2
\!+\!E)r^2\!-\!(\mu\!-\!1/2)^2]u
 &=& r^2 \Delta(r) v \nonumber \\
r^2 v^{\prime\prime}+rv^\prime \!+\![(k_r^2\!
-\!E)r^2\!-\!(\mu\!+\!1/2)^2]v
 &=& -r^2 \Delta(r) u,
\label{eq:BdG_eq}
\eary
with $k_r^2=k_f^2-k_z^2$. We set $2m\equiv 1$ for convenience and
restore it in the final expression of $M_v$.

In arriving at the mass formula we made a physically important assumption
that the dynamics of a vortex is {\it local in time}. From Eqs.\
(\ref{eq:eff_action}) and (\ref{eq:local_action}) one finds that such
assumption is valid
if the vortex motion is {\it slow}, i.e. $r_v (\tau)\!\approx\! r_v
(\tau')$,
on the relevant time scale $|\tau\!-\!\tau'|$. For situations where this
assumption is violated, one must work with the original, non-local action.

The eigenstates of the BdG equations are either {\it localized}
at the core
($|E|<\Delta_\infty$), or {\it extended} ($|E|>\Delta_\infty$), where
$\Del$ is a gap at infinity.
Accordingly the transition elements can be classified as {\it core-core},
{\it core-extended} ({\it extended-core}) and 
{\it extended-extended}. We examine the
three contributions  individually. The core solutions were first obtained
by Caroli {\it et al.}\cite{deGennes} and the extended states are
discussed by Cleary and by Bardeen {\it et al.}\cite{extended}

{\it Core-core}:
The core states are given by
\be
\left( \begin{array}{c}
              u(r) \\
              v(r)
       \end{array} \right)=
 {1\over2}\left( k_r \over \xi_0 \right)^{{1\over2}} \left(
   \begin{array}{c}
        J_{\mu\!-\!{1\over2}}(k_r r) \\
        J_{\mu\!+\!{1\over2}}(k_r r)
   \end{array} \right).
\label{eq:core_states}
\ee
The state is normalized over the radius $\xi_0$. The coherence
length $\xi_0$ is much larger than $k_r^{-1}$ in extreme type-II
superconductors considered here. Thus the derivatives of the radial
functions are
essentially those of the Bessel functions, allowing us to apply the usual
Bessel identities. We obtain
\be
-\langle \alpha |\nabla_v H_0 |\beta\rangle=
{k_r \epsilon_0 \over 2}(\hat{x}\mp i\hat{y})\delta_{\mua,\mub\pm1},
\label{eq:mat_ele_cc}
\ee
where $\eo=\Del/\kr\xi_0$ is the core level spacing and
$\ea=\mu_\alpha \eo$. From this we obtain the mass:
\bary
M_v^{cc} & = & m\left({{4k_f L}\over {3\pi}}\right)
\left(E_f \over \epsilon_0\right)
\sum_\mu |f(\eo \mu)\!-\!f(\eo\mu\!+\!\eo)|
\nonumber \\
&\approx&m\left({{4k_f L}\over {3\pi}}\right)
\left(E_f \over \epsilon_0\right)
[f(-\Del)\!-\!f(\Del)].
\label{eq:core_mass}
\eary
The factor $k_f L$ counts the number of electrons along the length of the
vortex $L$, given that the electron spacing is $k_f^{-1}$.
Since $E_f /\eo\sim (\kf \xi_0)^2$, the mass we obtained
above is comparable to the mass of the electrons displaced by a hollow
cylinder of radius $\xi_0$. The zero temperature limit of Eq.\
(\ref{eq:core_mass}) has been obtained previously\cite{mass,simanek}.
The temperature dependence is through the
difference of the Fermi distribution function at $\pm\Del$, which is
nearly constant for temperatures small compared with $T_c$. Close to
$T_c$, the gap itself is sensitive to temperature, and Eq.\
(\ref{eq:core_mass}) diverges as $\xi_0\sim(T_c \!-\!T)^{-1/2}$.

This somewhat unphysical divergence of mass hints to us that perhaps our
formula should not be taken too seriously near the critical temperature.
One possible reason is the transition element, Eq.\
(\ref{eq:mat_ele_cc}), which is based on Bessel function eigenstates. For
divering coherence length, it is doubtful that the expression
(\ref{eq:core_states}) can be taken seriously all the way to
$r\!=\!\xi_0\!\gg\! k_r^{-1}$.

The spectral function for the core is a delta peak at frequency
$\omega\!=\!\eo$. The (unapproximated) action due to the core-core
transitions is
\be
S_{eff}\!=\!
{{\kf^3 \eo^2 L}\over 12\pi}\int_0^\beta \! d\tau \int_{-\infty}^{\infty}
\! d\tau^\prime |r_v (\tau)\!\!-\!\!r_v (\tau')|^2 \!
e^{-\eo |\tau\!-\!\tau^\prime |}.
\label{eq:non_local}
\ee
When $\eo\!\sim\! (T_c\! -\! T)^{1/2}$ is small, the exponential factor
$e^{-\eo|\tau\!-\!\tau^\prime |}$ remains order unity for a wide range of
$|\tau\!\!-\!\!\tau^\prime |$. For a large time interval the vortex may
have moved a significant distance and not necessarily in a rectilinear
fashion. In this case we cannot make the approximation leading to Eq.\
(\ref{eq:local_action}). It still remains true, however, that Eq.\
(\ref{eq:non_local}) governs the dynamics of a vortex. A motion described
by Eq.\ (\ref{eq:non_local})  does not in general conserve energy. Hence a
{\it fast} motion whose characteristic time scale is much smaller than
$\eo^{-1}$ causes core quasiparticle transitions which dissipate energy.

{\it Core-extended}:
We choose $\alpha$ for the extended states, and $\beta$ for the core
states. For the extended states, there are two independent
solutions for a given energy, whose approximate forms are\cite{extended}
\bary
\left(\begin{array}{c}
            u(r) \\
            v(r)
       \end{array}
\right)_{E>\Del}&=&\sqrt{ \kp\over{2R}}
\left( \begin{array}{c}
          u(E) \\
          v(E)
       \end{array}
\right)\times J_{\mu\!-\!{1\over2}}(\kp r), \nonumber \\
 & &\sqrt{ \km\over{2R}}
\left( \begin{array}{c}
          v(E) \\
          u(E)
       \end{array}
\right)\times J_{\mu\!+\!{1\over2}}(\km r),
\label{eq:extended_states_Eplus}
\eary
where $u(E),v(E)\!=\!\{(E\!\pm\!
\sqrt{E^2 \!-\!\Delta^2 (r)})/2E\}^{1/2}$
and
$k_{\pm}\!=\!(k_r^2\pm\sqrt{E^2 \!-\!\Delta^2 (r)})^{1/2}$. The
normalization is over a large cylinder of radius $R$. The negative
energy solutions can be found from the substitution $(-v^*, u^*)$.
Employing Eq.\
(\ref{eq:m_eff2}) for the mass,
\be
|\langle \alpha|\nabla_v |\beta\rangle|^2
\approx
{\xi_0 \over{8R} }\left(\kr\!-\!{|\mua|\over \xi_0} \right)^2 \theta(\xi_0
\kr\!-\!|\mua|),
\ee
together with the constraint that the overlap integral is significant
only if
$\sqrt{\ea^2\!-\!\Del^2}<\Del$\cite{comment2}. 
The effective mass becomes ($x\equiv \beta\Del/2$)
\be
M^{ce}_v =m\left( {{ k_f L} \over {6\pi^2 }}\right)\left(E_f \over \eo
\right)
\int_{-1}^1 d\mu
{{(1\!-\!|\mu|)^{\kappa}}\over{1\!-\!\mu}}
{{\sinh[x(1\!-\!\mu)]} \over {\cosh[x]\cosh[x\mu]}}
\label{eq:ce_mass}
\ee
with $\kappa\approx 2.4$.

Replacing the integral with a characteristic value at $\mu\!=\!0$
immediately shows that the core-extended vortex mass has the temperature
dependence $\sim f(-\Del)\!-\!f(\Del)$, similar to the core-core mass. We
conclude that $M_v^{ce}$ is of the same order as $M_v^{cc}$, with the
similar temperature dependence. The extended-core transitions will give
the same mass as $M_v^{ce}$.

{\it Extended-extended}:
The matrix element appearing in Eq.\ (\ref{eq:m_eff1}) consists of two
parts. One involves the effect of ``bending" of the order parameter
$d\Delta(r)/dr$ and is only significant in the core region, $0\!<\! r\!
<\!\xi_0$.  The other transition comes from the phase gradient $\nabla
e^{-i\theta}$ which falls off as $1/r$, a long-ranged effect. One should
also distinguish the transitions between states which lie on the same side
of the Fermi surface, from those where a transition across the energy gap
is involved.  Their contributions to mass turn out to be very different. 

We consider the $\ea\eb>0$ case first.  Our approximate wavefunctions Eq.\
(\ref{eq:extended_states_Eplus})  and their negative energy counterparts
are supposed to be valid for $r$ not much bigger than $\xi_0$. As $r$ gets
well outside the core the wavefunction picks up an additional phase shift
not shown in Eq.\ (\ref{eq:extended_states_Eplus})\cite{extended}. 
Directly substituting our wavefunctions in the phase-gradient part of the
integral leads to very small values for $\ea\approx\eb$, where the
wavefunctions are likely to have the largest overlap. 
Hence a more accurate estimation of the wavefunction will not greatly
enhance the size of the phase-gradient integral. We will
thus assume that the integral can be restricted to the core region, where
one can put $\Delta(r)=\Del r/\xi_0$\cite{gygi}. With this substitution
one has $-\langle \alpha| \nabla_v H|\beta\rangle\approx 2\pi \Del
\xi_0^{-1} (\hat{x}\mp i\hat{y})\int_0^{\xi_0} r
u_{\alpha(\beta)}v_{\beta(\alpha)}dr$. We need to separately consider four
cases where $\alpha$ and $\beta$ are one of the extended eigenstates, and
sum the squared magnitude of the transition amplitude for each process. We
have
\be 
|\langle\alpha |\nabla_v H_0 |\beta\rangle|^2\approx 
{\Del^2 \over R^2}\theta(\xi_0 k_r \!-\!|\mua|)  
\label{eq:mat_ele_same_e} 
\ee 
provided
$(\ea^2\!-\!\Del^2)^{1/2}+(\eb^2\!-\!\Del^2)^{1/2}$ is on the order of
$\Del$ or less. Both states lie close to the energy gap, reflecting 
the fact that states far from the Fermi surface do not influence the
superconducting phenomenon.

We can compute the spectral function $J(\omega)$ according to Eq.\
(\ref{eq:spectralJ}). The sum over states is with respect to $k_z$,
$\mua=\mub\pm1$, and the radial wavenumbers for both states.
The integration measure is
\be
\sum_{k_z}\sum_{\mua \mub}\sum_{k_{\alpha} k_{\beta}}\approx
     {{LR^2 \xi_0^2}\over 2\pi^2}\int\!\!\int\!
{|\ea\eb| d\ea d\eb
\over{\sqrt{(\ea^2\!\!-\!\!\Del^2)(\eb^2\!\!-\!\!\Del^2)}}}.
\ee
When $\omega=0$, the integral is logarithmically divergent
because of $(E^2\!-\!\Del^2)^{-1}$ term. Finite $\omega$ effectively
serves as the lower cutoff, giving
\bary
J(\omega)&\approx& {{Lm^2 \xi_0 \Del^2 (\Del\!+\!\omega)}\over \pi}
\ln\left({\Del\over\omega}\right)\nonumber \\
&\times&[f(\Del)\!-\!f(\Del\!+\!\omega)].
\eary

In a previous calculation by two of the authors (Ao \& Zhu), it was
shown that a vortex experiences friction at non-zero temperature\cite{az}.
The friction coefficient
$\eta$ is obtained as the derivative of $J(\omega)$ at zero frequency
\cite{cl}:
\be
\eta=m\Del \left({{\kf L}\over 4\pi} \right)
{ {\beta \Del/2} \over{\cosh^2 [\beta\Del/2]}}
\ln\left({\Del \over\omega_c}\right)
\label{eq:friction}
\ee
with some lower frequency cutoff $\omega_c$ inserted by hand.
A constant friction  coefficient is a consequence of the {\it Ohmic}
character of the spectral function, $J(\omega)\sim \omega$ (up to a
logarithmic factor). In turn one finds from Eq.\ (\ref{eq:local_action})
that the mass in
this case is linearly divergent! A straightforward evaluation of the mass
gives
\be
M_v^{ee}=m \left({{k_f L} \over {2\pi^2}}\right){\Del\over \omega_c}
{\beta\Del/2 \over{\cosh^2 [\beta\Del/2]}}
\ln\left({\Del\over\omega_c}\right).
\label{eq:extended_mass}
\ee
As anticipated, the mass diverges linearly for a fixed temperature as
the low frequency cutoff tends to zero. For a fixed cutoff $\omega_c$
one gets exponentially vanishing mass in the zero temperature limit,
because only the gapped transitions across the Fermi surface contribute to
$J(\omega)$.

It was already mentioned  that the very concept of mass depends on the
motion the object executes. Here the mass is manifestly cutoff-dependent.
To understand the physical origin of the cutoff $\omega_c$ let us
go back to the original action, Eq.\ (\ref{eq:eff_action}). One can
divide the
integral over $\omega$ into two parts;
$0\!<\!\omega\!<\!\omega_c$ and $\omega_c\!<\!\omega\!<\infty$, where a
given motion of a vortex is considered slow/fast in comparison with
$\omega_c^{-1}$. For the fast spectral mode in
$\omega_c\!<\!\omega\!<\infty$ one can make the approximation
$r_v (\tau)\!-\!r_v (\tau')\!\approx\!(\tau\!-\!\tau' )
\dot{r}_v(\tau)$ while for the more slowly responding environment one
needs to keep the action as it is. If $\omega_c$ is much smaller than
$\Del$, $J(\omega)$ is nearly Ohmic  for
$\omega\! <\!\omega_c$, and the action becomes
\bary
S_{eff}&\approx &{1\over2}\int_0^\beta d\tau\!
 \left\{\int_{\omega_c}^\infty d\omega {4J(\omega)  
 \over{\pi\omega^3}}\right\}
\left({{d\rv}\over d\tau}\right)^2\!
 \nonumber \\
  &+& {\eta\over {2\pi}}\int_0^\beta d\tau \int_{-\infty}^{\infty}
d\tau^\prime \left|
{{r_v (\tau)\!\!-\!\!r_v (\tau')}\over 
{\tau\!-\!\tau^\prime}} \right|^2.
\eary
This is the standard Caldeira-Leggett model of a Brownian
particle\cite{cl}, where
the mass and friction are given by Eqs.\ (\ref{eq:extended_mass})
and (\ref{eq:friction}).
The friction coefficient is related to the mass by
$M^{ee}_v=2\eta/\pi\omega_c$. The classical equation of motion becomes
\be
M_v^{ee} \dot{v}+\eta v=\eta
\left( {2\over{\pi\omega_c}}\dot{v}+ v\right).
\ee
Since the scale $\omega_c$ is itself set by the motion, it is reasonable
to assume that $|\dot{v}|\approx |\omega_c\, v|$, indicating that the
effects of inertia and drag are always of the same order regardless of the
characteristic time scale of the motion. 

Transitions across the Fermi level obviously do not contribute to
friction, since the energy difference of the initial and final states
is greater than $2\Del$. We also expect that the mass contribution from
this process is finite. In estimating the overlap integral we ignored the
integration in the far region, $\xi_0\!<\!r\!<\!R$, because the
phase shifts for states $\ea$ and $\eb$, $\ea+\eb\approx 0$, should
generically be different. Integrating over the core region gives
\be
|\langle\alpha|\nabla_v H_0 |\beta\rangle|^2\approx
{\Del^2 \over R^2 }\left(2- {\Del^2 \over \ea^2}\right)
\ee
provided $k_r \xi_0 >|\mua|\approx|\mub|$. It is similar in
magnitude to Eq.\ (\ref{eq:mat_ele_same_e}) for the $\ea\eb\!>\!0$ case.
However the energy denominator is now of order
$(\ea-\eb)^3 \!\approx\! \Del^{3}$ and free of divergence. The mass is
given by $m(k_f L)[f(-\Del)\!-\!f(\Del)]$ up to a numerical factor of
order unity.
While the temperature dependence is similar to $M_v^{cc}$ and
$M_v^{ce}$ numerically it is much smaller than both.

A vortex is not by itself assigned an intrinsic mass. Rather it is a
consequence of interaction with the surrounding electrons.
The motion of a vortex causes transitions between various BdG eigenstates
and accrues mass as a consequence.

We showed that both $M_v^{cc}$ and $M_v^{ce(ec)}$ are comparable to the
mass of electrons displaced by the hollow core in the $T=0$ limit, and has
a temperature dependence $\sim\xi^2_0 (T)  [f(-\Del)\!-\!f(\Del)]$.  The
transition between extended states with small energy differences leads to
a frequency-dependent mass and a friction at finite temperature. Both
quantities vanish at zero temperature exponentially. Considering extended
states alone, one concludes that the vortex motion is dissipative at
finite $T$ even in the absence of impurities. In the low temperature range
where $M_v^{cc}$ and $M_v^{ce}$ are large and the friction is small, the
vortex motion is Newtonian (mass times acceleration equals force), with a
large mass. Experimental indications may have already existed \cite{exp}. 
If the characteristic time scale of the motion is much smaller than the
inverse core level spacing, however, the core level transitions can lead
to dissipation also.

P. Ao and X.-M. Zhu were supported by the Swedish Natural Science Research
Council (NFR).


\widetext


\begin{references}
\bibitem{comment} We consider a rectilinear vortex of
length $L$ along the $z$ direction.
\bibitem{comment2}The constraint follows from integrating over a product
of two Bessel functions whose arguments are $k_r r$ and $k_{\pm}r$,
respectively, over the coherence length, since $|k_r \!-\!k_{\pm}|$   
differing by more than $\xi_0^{-1}$ will make the integral very small.
\end{references}
\end{document}